# An Intelligent and Low-Cost Eye-Tracking System for Motorized Wheelchair Control

Mahmoud Dahmani [1], Muhammad E. H. Chowdhury [2], Amith Khandakar [2], Tawsifur Rahman[3], Khaled Al-Jayyousi [2], Abdalla Hefny [2] and Serkan Kiranyaz [2,*]

1   School of Engineering, University of Maryland, College Park, MD 20742, USA; dahmani@umd.edu
2   Department of Electrical Engineering, College of Engineering, Qatar University, Doha 2713, Qatar; mchowdhury@qu.edu.qa; amitk@qu.edu.qa; ka1403027@student.qu.edu.qa; ah1406234@student.qu.edu.qa
3   Department of Biomedical Physics and Technology, University of Dhaka, Dhaka 1000, Bangladesh; tawsifurrahman1426@gmail.com
*   Correspondence: mkiranyan@qu.edu.qa; Tel.: +974 3063 5600

**Abstract:** In the 34 developed and 156 developing countries, there are ~132 million disabled people who need a wheelchair, constituting 1.86% of the world population. Moreover, there are millions of people suffering from diseases related to motor disabilities, which cause inability to produce controlled movement in any of the limbs or even head. This paper proposes a system to aid people with motor disabilities by restoring their ability to move effectively and effortlessly without having to rely on others utilizing an eye-controlled electric wheelchair. The system input is images of the user's eye that are processed to estimate the gaze direction and the wheelchair was moved accordingly. To accomplish such a feat, four user-specific methods were developed, implemented, and tested; all of which were based on a benchmark database created by the authors. The first three techniques were automatic, employ correlation, and were variants of template matching, whereas the last one uses convolutional neural networks (CNNs). Different metrics to quantitatively evaluate the performance of each algorithm in terms of accuracy and latency were computed and overall comparison is presented. CNN exhibited the best performance (i.e., 99.3% classification accuracy), and thus it was the model of choice for the gaze estimator, which commands the wheelchair motion. The system was evaluated carefully on eight subjects achieving 99% accuracy in changing illumination conditions outdoor and indoor. This required modifying a motorized wheelchair to adapt it to the predictions output by the gaze estimation algorithm. The wheelchair control can bypass any decision made by the gaze estimator and immediately halt its motion with the help of an array of proximity sensors, if the measured distance goes below a well-defined safety margin. This work not only empowers any immobile wheelchair user, but also provides low-cost tools for the organization assisting wheelchair users.

## 1. Introduction

The human eye is considered to be an intuitive way of interpreting human communication and interaction that can be exploited to process information related to the surrounding observation and respond accordingly. Due to several diseases like complete paralysis, multiple sclerosis, locked-in syndrome, muscular dystrophy, arthritis, Parkinson's, and spinal cord injury, the person's physiological abilities are severely restricted from producing controlled movement in any of the limbs or even the head, noting that there are about 132 million disabled people who need a wheelchair, and only 22% of them have access to one [1]. They cannot even use a technically advanced wheelchair. Thus, it is very important to investigate novel eye detection and tracking methods that can enhance human–computer interaction, and improve the living standard of these disable people.

Research for eye tracking techniques has been progressively implemented in many applications, such as driving fatigue-warning systems [2,3], mental health monitoring [4,5], eye-tracking controlled wheelchair [6,7], and other human–computer interface systems. However, there are several constraints such as reliable real-time performance, high accuracy, availability of components, and having a portable and non-intrusive system [8–10]. It is also crucial to achieve higher system robustness against encountered challenges, such as changing light conditions, physical eye appearance, surrounding eye features, and reflections of eye-glasses. Several related works have proposed eye-controlled wheelchair systems; however, these rarely address the constraints of the system's software performance, physical, and surrounding challenges beyond the system, novelty of algorithms, and user's comfort and safety altogether.

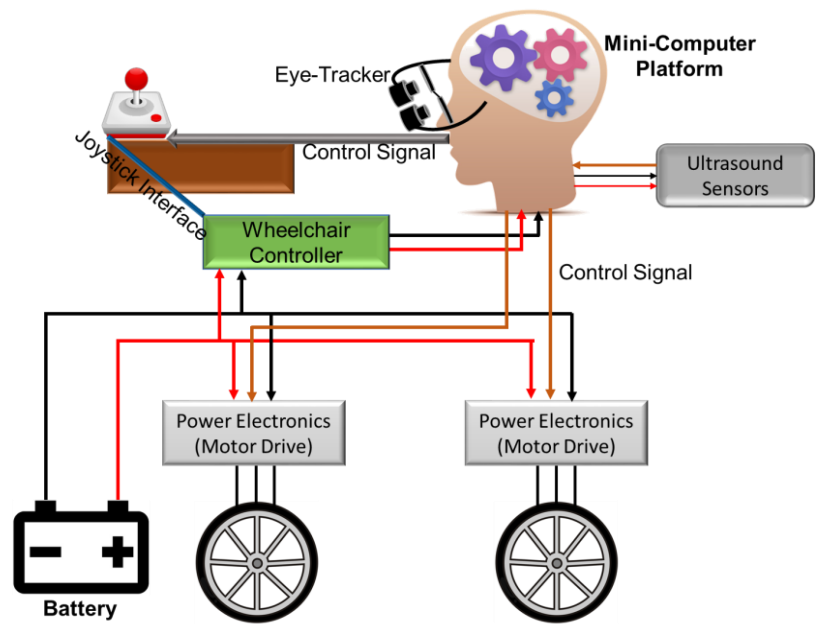

**Figure 1.** Block representation of the proposed systems for pupil based intelligent eye-tracking motorized wheelchair control.

Furthermore, the Convolutional Neural Network (CNNs) is a state-of-the-art and powerful tool that enables solving computationally and data-intensive problems. CNN is pioneering in a wide spectrum of applications in the context of object classification, speech recognition, and natural language processing and even wheelchair control [11–13], a more detailed literature review will be discussed in the later sections. However, the paper lacks high accuracy, real-time application, and does not provide the details of such a design, which could be useful for further improvement. All of the above shortcomings are addressed in the current paper.

In this paper, we propose a low-cost and robust real-time eye-controlled wheelchair prototype using novel CNN methods for different surrounding conditions. The proposed system comprises of two subsystems: sensor subsystem and intelligent signal processing, and decision-making and wheelchair control subsystem as illustrated in Figure 1. The sensor subsystem was designed using an eye-tracking device and ultrasound sensors, which were interfaced to the intelligent data processing and decision-making module. The motor control module was already available in the powered wheelchair; only control signals based on the eye-tracker needs to be delivered to the microcontroller of the original wheelchair joystick bypassing the mechanical joystick input. An array of ultrasound sensors was used to stop the wheelchair in case of emergency. The proposed system can steer through a crowded place faster and with fewer mistakes than with current technologies that track eye movements. The safety provision, ensured by arrays of ultrasound sensors, helps the wheelchair to steer through a congested place safely. Accordingly, the proposed system can help most of the disabled people with spinal cord injury. Furthermore, as the proposed system is targeted to use inexpensive hardware and open source software platform, it can even be utilized to modify non-motorized wheelchairs to produce a very economical motorized wheelchair solution for third-world countries.

The rest of this paper is outlined as follows. Section 2 provides background and reviews the relevant literature on state-of-the-art Eye-Tracking methods, existing eye-controlled wheelchair systems, and Convolutional Neural Networks (CNNs) for Eye Tracking. Section 3 is the Methodology section, which discusses the design of the various blocks of the work along with the details of the machine learning algorithm. Section 4 provides the details of the implementation along with the modifications done in the hardware. Section 5 summarizes the results and performance of the implemented system. Finally, the conclusion is stated in Section 6.

## 2. Background and Related Works

Previous studies are explored in this paper within three contexts in order to investigate all major related aspects as well as to cover the relevant literature as much as possible. The aspects are state-of-the-art methods for eye tracking, existing eye-controlled wheelchair systems, and other convolutional neural network (CNN)-based works for eye tracking application.

### 2.1. State-Of-The-Art Eye Tracking Methods

Generally, there are two different methods investigated widely for eye tracking: Video-based systems and Electrooculography (EOG)-based systems (Figure 2). A video-based system consists mainly of a camera placed at a distance to the user (remote), or attached to the user's head (head-mounted), and a computer for data processing [14,15]. However, the main challenge in remote eye tracking is robust face and eye detection [16,17]. The cameras can be visible-light cameras, referred to as Videooculography (VOG) [14], as examples proposed in [9,18–20], or infrared-illumination cameras such as in [21,22], where infrared (IR) corneal reflection was extracted. Based on near IR illumination, researchers in [17] investigated six state-of-the-art eye detection and tracking algorithms: Ellipse selector (ElSe) [23], Exclusive Curve Selector (ExCuSe) [24], Pupil Labs [25], SET [26], Starburst [27], and Swirski [28], and compared them against each other on large four datasets with an overall 225,569 public labeled eye images of frequently changing sources of noise.

Commercial eye-tracking systems are still very expensive using proprietary tracking algorithms that are not commercially deployed to any powered wheelchair. Note that although pupil tracking is a widely used tool, it is still hard to achieve high-speed tracking with high-quality images, particularly in a binocular system [29].

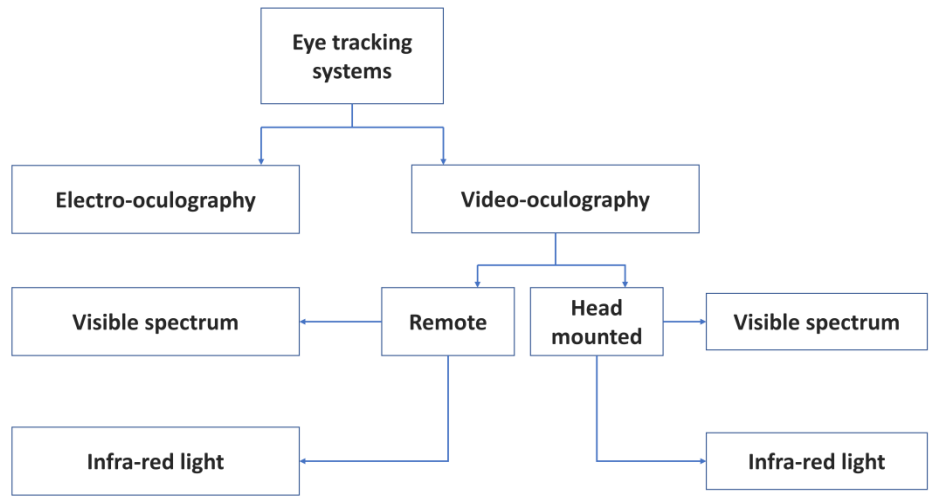

**Figure 2.** Different eye tracking techniques.

*2.2. Existing Eye-Controlled Wheelchair Systems*

User's voice [30] and facial expression [31]-based wheelchair control systems were explored by different groups. However, voice control is laborious for the user and sound waves interference or noisy environment distractions can be introduced to the system establishing undesired commands. Facial expression control, on the other hand, is not helpful to all users, especially those who suffer from restrictions in facial expressions due to diseases like facial paralysis. Moreover, classification of facial expressions is more challenging than the eye-controlled system, where only the eye is targeted.

Eye tracking techniques have been previously employed to serve wheelchair systems for the disabled people. An eye-controlled wheelchair prototype was developed in [32] using an infrared (IR) camera fitted with LEDs for automatic illumination adjustment during illumination changes. A flow of image processing techniques was deployed for gaze detection beginning with eye detection, pupil location detection using pupil knowledge, and then converting the pupil location into the user's gaze based on a simple eye model. This was finally converted into a wheelchair command using a threshold angle. Although the algorithm had a fast processing time and robustness against changing circumstances, the used threshold was suitable only to a specific illumination condition enabling automatic illumination adjustment by the camera. This degrades significantly when hit by a strong illumination such as sunlight. In addition, the conceptual key control was quite inconvenient for the user, as the chair stops when the user blinks and when the gaze is deviated from the direction key unless the user looks upwards for free head movement. The user also has to look downwards for forward movement, which is impractical.

In [33], another eye-controlled wheelchair is proposed by processing the images of a head-mounted camera. Gaussian filtering was implemented for removing Gaussian noise from the image. A threshold was then employed for producing a binary image, and erosion followed by dilation was applied for removing white noise. The wheelchair moves in three directions (left, right, and forward) depending on the relative iris position, and starts and stops by blinking for 2 seconds. Although the proposed techniques were simple in implementation, the evaluation parameters of the system were not reported. In addition, the system's performance during pupil transition from one direction to another was not discussed.

Apart from depending on interfaces like joystick control, head control, or sip-puff control [34], the optical-type eye tracking system that controls a powered wheelchair by translating the user's eye movement on the screen positions was used by several researchers, which are reported below. The eye image was divided into nine blocks of three columns and three rows, and depending on the location of the pupil's center, the output of the algorithm was an electrical signal to control the wheelchair's movement to left, right, and straight directions [34]. The system evaluation parameters like response speed, accuracy, and changing illumination conditions were not reported. In addition, safety parameters of the wheelchair's movement, such as ultrasound or IR sensors for obstacles detection, were not discussed.

Another wheelchair control system has been proposed in [35], where positions of the eye pupil were tracked by employing image processing techniques using a Raspberry-Pi board and a motor drive to steer the chair to left, right, or forward directions. The open computer vision (OpenCV) library was used for image processing functions, where the HAAR cascade algorithm was used for face and eye detection, Canny edge was used for edges detection, and Hough Transform methods were used for circle detection to identify the border of the eye's pupil. The eye pupil's center is located depending on the average of two corner points obtained from a corner detection method. The pupil was tracked by measuring the distance between the average point and the eye circle's center point, where the minimum distance indicated that the pupil was at left, and the maximum indicated the eye had moved to the right. If there was no movement, the eye would be in the middle position, and the chair would be moving forward. Eye blinking was needed to start a directional operation, and the system was activated or deactivated when the eye was closed for 3 seconds. Although the visual outputs of the system were provided, the system was not quantitatively assessed, and no rigid evaluation scheme was shown. Therefore, accuracy, response latency, and speed for instance are unknown.

On the other hand, Electrooculography (EOG) is a camera-independent method for gaze tracking, and generally, it requires lower response time and operating power than the video-based methods [10]. In EOG, electrodes are placed on the skin at different positions around the eyes along with a reference electrode, known as ground electrode, placed on the subject's forehead. The eye is modeled as an electric dipole, where the negative pole is at the retina and the positive pole is at the cornea. When the eyes are in their origin state, the electrodes measure a steady electric potential, but when an eye movement occurs, the dipole's orientation changes making a change in the electric corneal–retinal potential that can be measured.

An EOG-based eye controlled wheelchair with an attached on-board microcontroller was proposed in [36] for disabled people. Acquired biosignals were amplified, noise-filtered, and fed to a microcontroller, where the input values for each of the movement (left, right, up, and down) or stationary conditions were given, and the wheelchair movement in the respective direction was performed according to the corresponding voltage values from the EOG signals. The EOG-based system was cost effective, independent from changing light conditions, and containing lightweight signal processing with reasonable functional feasibility [10]; however, the system was not yet fully developed for commercial use because of the electrodes used for signal acquisition. Moreover, it was also restricted for particular horizontal and vertical movements of the eye; it did not respond effectively for oblique rotation of the eye.

Another EOG-guided wheelchair system was proposed in [37] using the Tangent Bug Algorithm. The microcontroller identified the target point direction and distance by calculating the gaze angle that the user was gazing at. Gaze angle and blinks were measured, and used as inputs for the controlling method. The user only asked to look at the desired destination and blink to give the signal to the controlling unit for starting navigation. After that, the wheelchair calculated the desired target position and distance from the measured gaze angle. Finally, the wheelchair moved towards the destination in a straight line and go around obstacles when detected by sensors. Overall, EOG-based systems are largely exposed to signal noise, drifting, and artifacts that affect EOG signal acquisition. This is due to the interference of noise from residential power lines, electrodes, or circuitry [14]. In addition, with placing electrodes at certain distances around the eye, EOG-based systems are considered to be impractical for everyday use.

There are some recent works on using commercially available sensors for eye controlled powered wheelchair for Amyotrophic lateral sclerosis (ALS) patients [38]. A similar work using commercial brain–computer interface and eye tracker devices was done in [6].

*2.3. Convolutional Neural Networks (CNNs) for Eye Tracking*

CNN is a pioneer in a wide spectrum of applications in the context of object classification; however, only a few previous studies have presented CNNs for the specific task of real-time eye gaze classification to control a wheelchair system as in the current paper. Some examples for CNNs employed for eye gaze classification application are discussed below.

The authors of [39] proposed an eye-tracking algorithm that can be embedded in mobiles and tablets. A large training dataset of almost 2.5 M frames was collected via crowdsourcing of over 1450 subjects, and used for training the designed deep end-to-end CNN. Initially, the face image was used as original image, and the images of the eyes were used as inputs to the model. For real-time practical application, dark knowledge was then applied to reduce the computation time and model complexity by learning a smaller network that achieves a similar performance running at 10–15 frames per second (FPS). The model's performance increases significantly when calibration was done, and when there was variability in the collected data with higher number of subjects rather than higher number of images per subject, but not neglecting the importance of the second. Although the model has achieved robustness in eye detection, the FPS rate should not be less than 20–25 FPS for reliable real-time performance; a rate of which the error would increase significantly in this case if was not addressed.

The authors of [40] proposed a real-time framework for classifying eye gaze direction applying CNNs, and using low-cost off the shelf webcams. Initially, the face region was localized using a modified version of the Viola–Jones algorithm. The eye region was then obtained using two different methods: the first one was geometrically from the face bounding box, and the second (which showed better performance) was localizing facial landmarks through a facial landmark detector to find the eye corners and other fiducial points. The eye region was then catered to the classification stage, where classes of eye access cues are predicted and classified using CNN into seven classes (center, upright, up left, right, left, downright, and down left), where three classes (left, right, and center) showed higher accuracy rates. The algorithm's evaluation was performed on two equal data 50% subsets of testing and training, where CNNs are trained for left and right eyes separately, but eye accessing cues (EAC) accuracy was improved when combining information from both eyes (98%). The algorithm has achieved an average rate of 24 FPS.

The authors of [12] implemented an eye-controlled wheelchair using eye movement captured using webcam in front of the user and using Keras deep learning pre-trained VGG-16 model. The authors also discuss the benefits of the project working for people with glasses. Details of its real-time implementation, in terms of FPS, was not mentioned in the paper.

## 3. Methodology

Considering the pros-and-cons of the previous works, we propose an eye-controlled wheelchair system running at real-time (30 FPS) using CNNs for eye gaze classification, and made up of several low-cost, lightweight controller subsystems for wheelchair. The system takes the input images from an IR camera attached to a simple headset, providing comfort and convenient movement control to the user with complete free-eye movement. Ultrasonic sensors were mounted to avoid collisions with any obstacles. The system classifies the targeted direction based on a robust CNN algorithm implemented on Intel NUC (a mini-computer with I7-5600U) at 2.4 GHz (2 central processing units (CPUs)) and 8 GB random access memory (RAM) using C++ in Microsoft Visual Studio 2017 (64 bit). Although it is not a graphics processing unit (GPU)-optimized implementation, multiprocessing with a shared memory was attained by deploying the Intel® OpenMP (Open Multi-Processing) application programming interface (API).

A block diagram for the eye-controlled wheelchair is shown in Figure 3. The diagram shows different steps, starting from capturing a new image, until the appropriate command, which is given to the wheelchair.

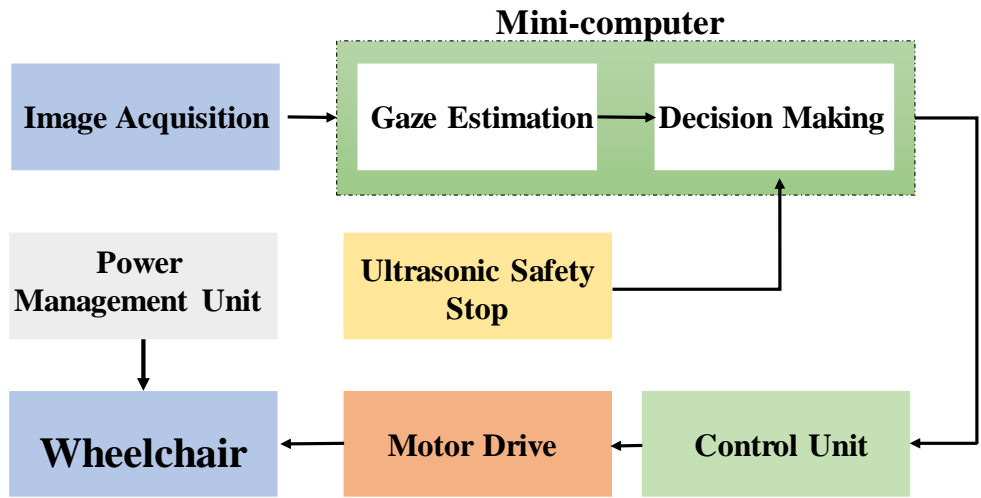

**Figure 3.** Eye Controlled Wheelchair block diagram.

The wheelchair is primarily controlled through the eye movements that are translated into commands to the chair's motor drives. This is achieved through a gaze detection algorithm implemented on the minicomputer. Yet, a secondary commanding system was added to the design for the means of safety of the user. This safety system is ultrasonic-based that can stop the wheelchair in cases of emergency, suddenly appearing objects, unawareness of the user, etc.

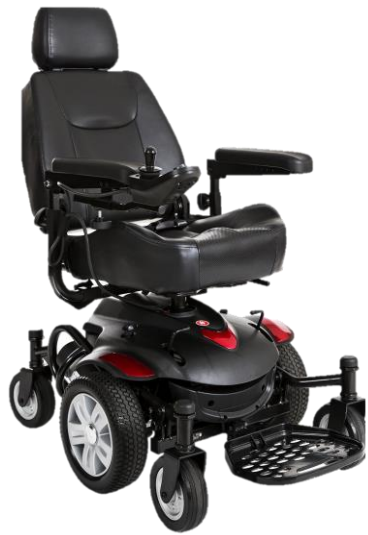

**Figure 4.** Titan x16 wheelchair.

The minicomputer should make the decision of whether the wheelchair should move next mainly according to the gaze direction, or to stop the wheelchair if the safety system is activated. In either case, a command is sent to the control unit, which in turn produces the corresponding command code to send to the wheelchair controller to drive the motor to move the wheelchair in the respective direction.

The Titan x16 power wheelchair was converted to an eye-tracking motorized wheelchair (Figure 4). It is originally equipped with a joystick placed at the right arm of the chair. An eye-tracking controller (image acquisition system and mini-computer) was connected to the electronic control system beside the joystick-based controller. Furthermore, the functionality of the main joystick was not altered; the new control system can be superseded by the original control system.

Based on this overview, the system implementation can be divided into the following steps; design and implementation of an image acquisition mechanism, implementation of the gaze estimation algorithm, design and implementation of the ultrasonic safety system, and finally modifying the joystick controller. Each of these parts is discussed separately as below.

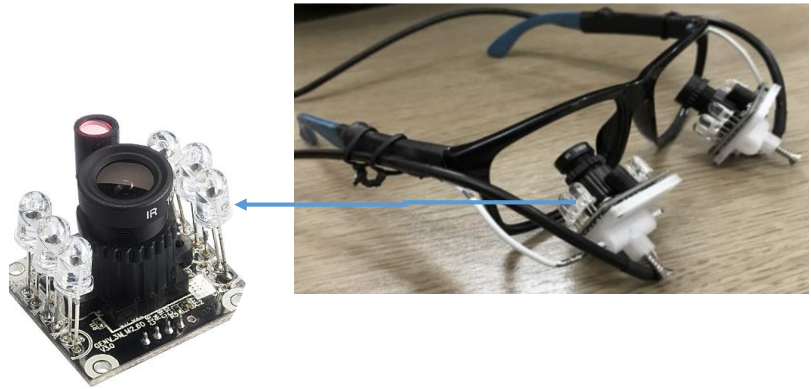

**Figure 5.** Image acquisition system using IR camera.

*3.1. Image Acquisition Frame Design*

The concept of the design for an eye-frame was first visualized using Fusion 360 software to validate its user friendliness. Two infrared (IR) [41] cameras were mounted on small platforms supported by flexible arms below eye level. Flexible arms were used to allow position adjustment of the two cameras for different users. The cameras were positioned at a distance that allowed the full eye movement range to be recorded, as seen in Figure 5. The reason for using the IR cameras was primary because (i) the acquired images were almost the same under different lighting conditions and (ii) the user was not irritated by the light emitted from the IR LEDs.

*3.2. Gaze Estimation Algorithm*

Accuracy, speed, and robustness to variation in illumination are the key parameters of any gaze estimation algorithm. The work presented a comparison between different algorithms of gaze estimation using these three parameters. Numerous gaze estimation algorithms were implemented by the researchers; nevertheless, the focus of this work is on those algorithms which are user-specific. Figure 6 shows a set of such algorithms. The user specificity means that the algorithms initially trained for a particular user using sample images and then use the trained algorithm to estimate gaze directions. Those four algorithms with blue color in Figure 6 are those tested in this work.

To facilitate the disabled user to have full mobility for the wheelchair, at least three commands are required: forward, right, and left. Furthermore, more than two commands are needed to start and stop the process of moving the chair. It was decided to use left eye winking to start acquisition of images and moving the chair, and another left eye winking to stop acquisition. Correspondingly, for each eye, four states are required: Right Gazing, Forward Gazing, Left Gazing, and Closing the eye.

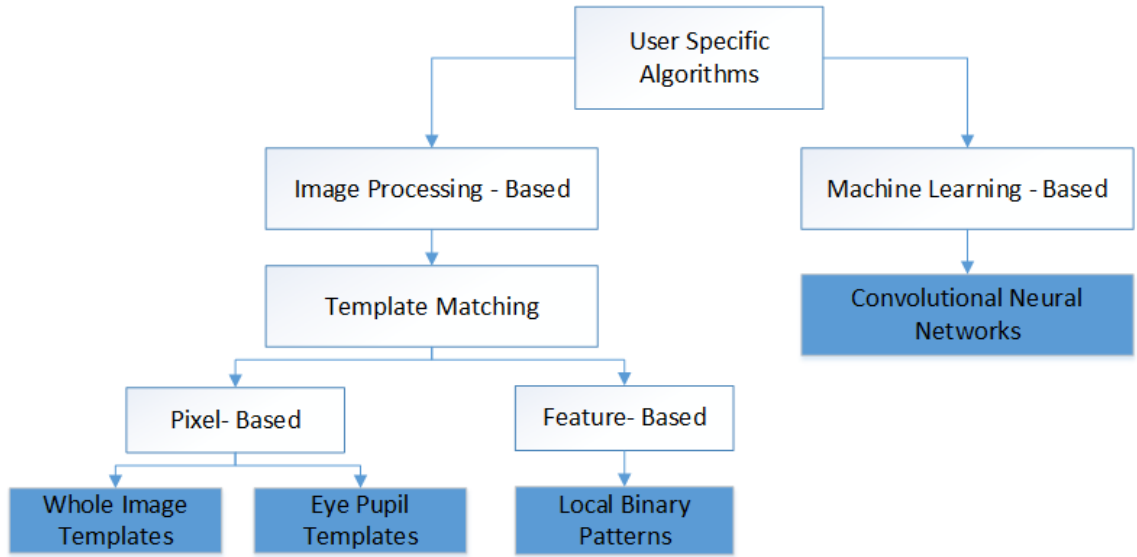

**Figure 6.** Classification of algorithms for gaze estimation.

The following subsection discusses different algorithms tested in this work.

(a) Template Matching

The process of template matching between a given patch image (known as a template T) and a search image S, simply involves finding the degree of similarity between the two images. The template images are usually smaller in size (lower resolution). The simplest way to find the similarity is the Full Search (FS) method [25], where the template image moves (slides) over the search image, and for each a new position of T, the degree of similarity of the pixels of both images is calculated. For each new position of T, a mathematical equation was used to measure the degree of matching, which was a correlation operation as represented in Equation 1, where T stands for template, S for search image, and the prime superscript represents the X or Y coordinates for the template image.

$$R(x,y) = \sum_{x'y'} T(x'.y').S(x+x'.y+y') \qquad (1)$$

Template Matching Using Whole Image Templates

To use template matching for gaze estimation, a set of search images should be available for each user (this set of images were collected in a calibration phase). Four search images were used: three for the three different gaze positions and one for the closed eye. The template image was the image with unknown gaze direction (but certainly, it was one of the four classes). For each new template image, the correlation is done four times (one for each class).

It is worth mentioning that the search and template images should be acquired with the same resolution. The template matching was not sliding, rather it was using a single correlation. The correlation process returned a correlation coefficient for each search image, and the search image with the highest correlation was selected as the class for the template image. Figure 7 shows how the gaze estimation algorithm using equal-sized images was applied.

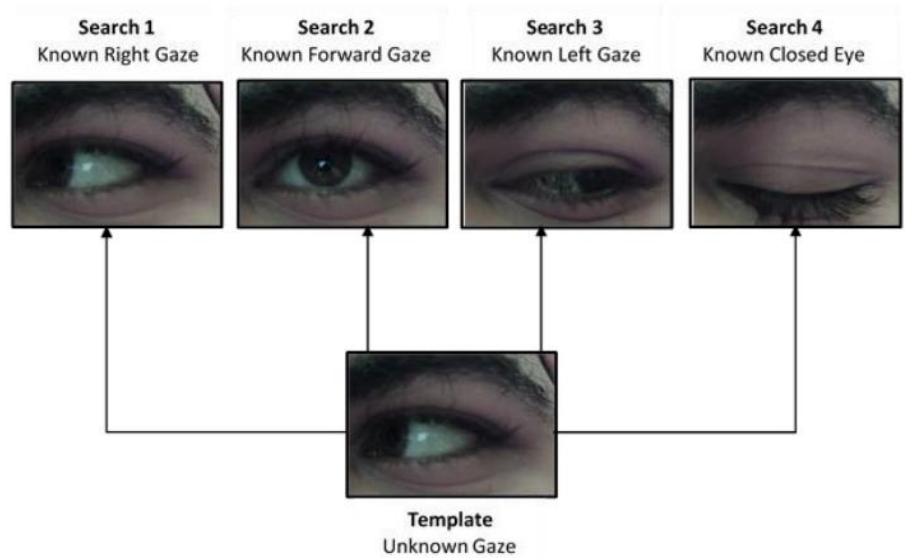

**Figure 7.** Gaze estimation using template matching.

The process shown in Figure 7 requires that the template and search image should be acquired in the same illumination condition. However, the search images should be collected only once at the beginning and should not be changed, but the template image was subjected to such changes in the lighting conditions. Accordingly, to manage the changes in illumination, Histogram Equalization (HE) was used. Histogram Equalization changes the probability distribution of the image to a uniform distribution to improve the contrast.

Template Matching Using Eye Pupil Templates

The second template matching based method uses correlation, but with a slight variation. In this method, the template is an image that contains only the pupil of the user. It was extracted in the calibration phase while the user was gazing forward; this gaze direction allowed for easier and simpler pupil extraction. The iris could have been used as a template; however, the pupil was a better choice because it always appears in all gaze directions (assuming an open eye). This is evident in Figure 7, where the whole iris does not appear in the case of right and left gazes. It is notable that in this template matching approach, only an image of the pupil was required as the template. The newly acquired images (with unknown pupil position) were considered as the search images.

The correlation was applied for each new search image by sliding the pupil on all possible positions of the search image with the template image to locate the position of the pupil on the search image that was maximally matched to the template. This method has two main advantages over the previous method. First, the correlation was done only once for each new image (instead of four times). The second merit was eliminating the constraint of having only few number of known gaze directions. This method can locate the pupil position at any location in the eye, not only forward, right, and left.

Feature-Based Template Matching

Another template matching technique was used to correlate feature values rather than raw pixel intensities. Local Binary Patterns (LBP) represent a statistical approach to analyze image texture [42]. The term "local" is used because texture analysis is done for each pixel with pixels in the neighborhood. The image was divided into cells, 3 × 3 pixels each. For each cell, the center pixel is surrounded by eight pixels. Simply stated, the value of the center pixel was subtracted from the surrounding pixels. Let $x$ denote the difference, then the output of each operation s($x$) was a zero or one, depending on the following thresholding.

$$s(x) = \begin{cases} 1, x \geq 0 \\ 0, x < 0 \end{cases} \tag{2}$$

Starting from the upper left pixel, the outputs were concatenated into a string of binary digits. The string was then converted to the corresponding decimal representation.

This decimal number was saved in the position of the center pixel in a new image. Therefore, the LBP operator returned a new image (matrix) that contains the extracted texture features. The template matching was done between the LBP outputs of the source and template images. Although applying this operation requires more time than directly matching the raw pixels, the advantage provided by LBP is the robustness to monotonic variations in pixels intensities due to changes in illumination conditions [43]. Thus, applying the LBP operator eliminates the need of equalizing the histogram.

(b) Networks Architecture and Parameters

In compliance with the notion of a user-specific approach, Convolutional Neural Networks (CNNs) were proposed as another alternative to classify gaze directions in a fast and accurate manner. Like other template matching-based methods, a calibration process should to be carried out to generate a labeled dataset that is required to train the supervised network. This is required to be performed only once for the first time the user utilizes the wheelchair. As a result, relatively small user-specific data, which can be acquired in less than 5 minutes at 30 frames/sec, were employed to train a dedicated CNN for each user. Therefore, the training time will be short and the trained classifier was well suited for real-time eye tracking by predicting the probabilities of the input image (i.e., the user's eye) being one of four classes: right, forward, left, and closed.

This proposed solution was data-driven approach and comprises the following stages.
1. Network Architecture Selection
2. Data Preprocessing
3. Loss function
4. Training algorithm selection
5. Hyperparameters setting

Networks Architecture Selection

The convolutional neural network (CNN) was selected as a network model because it produces state-of-the-art recognition results according to the literature. The main merit of CNNs is the fact that they combine the two major blocks of feature extraction and classification into a unified learner, as shown in Figure 8.

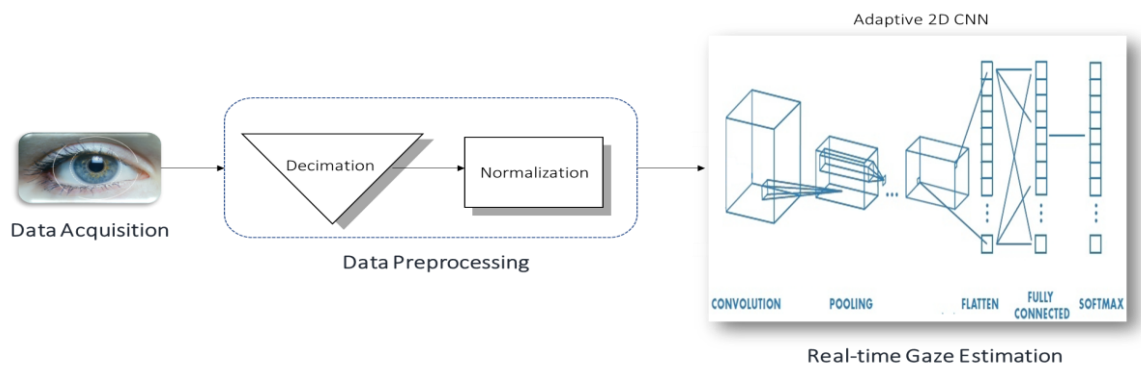

**Figure 8.** Block diagram of convolutional neural network (CNN).

In fact, this advantage is very convenient to this application for two reasons: First, CNNs learn directly on raw data (i.e., pixel intensities of the eye images), and thus eliminate the need for manual feature extraction. More importantly, the automation of feature extraction by means of training goes hand in hand with the philosophy of user-specificity because features that best characterize the eye pupil are learned for each individual user. By contrast, the previously discussed template-matching approach is not truly user-specific, as it operates on features that are fixed and handcrafted for all users, which does not necessarily yield the optimal representation of the input image.

Architecture wise, CNNs are simply feedforward artificial neural networks (ANNs) with two constraints:
1. Neurons in the same filter are only connected to local patches of the image to preserve spatial structure.
2. Their weights are shared to reduce the total number of the model's parameters.

A CNN consists of three building blocks:
1. Convolution layer to learn features.
2. Pooling (subsampling) layer to reduce the dimensionality the activation maps.
3. Fully-connected layer to equip the network with classification capabilities.

The architecture overview is illustrated in Figure 9.

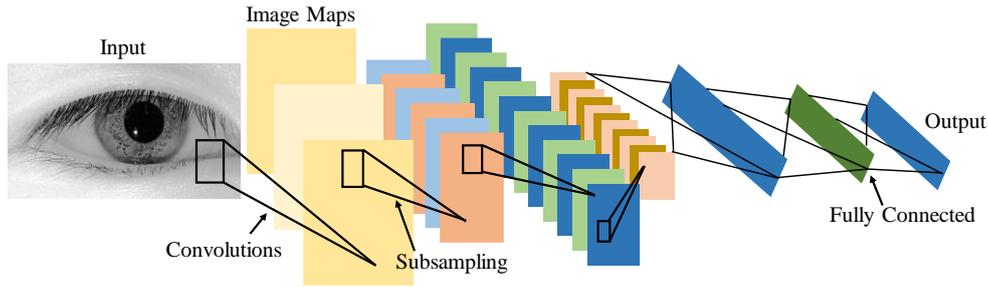

**Figure 9.** Convolutional Neural Network (CNN) architecture.

Data Preprocessing

Before feeding input images to the network, they were decimated to 64 × 64 images regardless of their original sizes. This reduced the time for the forward propagation of a single image when the CNN is deployed for real-time classification. Then, they were zero-centered by subtracting the mean image and normalized to unit variance by dividing over the standard deviation:

$$I_N(x,y) = \frac{I(x,y) - I_{mean}}{\sigma_I} \quad (3)$$

where $I$ is the original image; the $I_N$ is the preprocessed one; and $I_{mean}$ and $\sigma_I$ are the mean and standard deviation of the original image, respectively.

The preprocessing stage will help reduce the convergence time of the training algorithm.

Loss Function

The mean squared error (MSE) is the error function chosen to quantify how good the classifier is by measuring its fitness to the data:

$$E(weights, biases) = \sum_{i=1}^{N_c} (y_i - t_i)^2 \quad (4)$$

where $E$ is the loss function; $y_i$ and $t_i$ are the predicted and the actual scores of the $i^{th}$ class, respectively; and $N_c$ denotes the number of classes which is 4 in this classification problem.

Training Algorithm Selection

Learning is actually an optimization problem where the loss function is the objective function and the output of the training algorithm is the network's parameters (i.e., weights and biases) that minimize the objective.

Stochastic gradient descent with adaptive learning rate (adaptive backpropagation) is the employed optimization algorithm to train the CNN. Backpropagation involves iteratively computing the gradient of the loss function (the vector of the derivatives of the loss function with respect to each weight and bias) to use it for updating all of the model's parameters:

$$w_{ki}^l(t+1) = w_{ki}^l(t) - \varepsilon \frac{\delta E}{\delta w_{ki}^l} \quad (5)$$

$$b_{ki}^l(t+1) = b_{ki}^l(t) - \varepsilon \frac{\delta E}{\delta b_{ki}^l} \qquad (6)$$

where $w_{ki}^l$ is the weight of the connection between the *kth* neuron in the *lth* layer and the *ith* neuron in the *(l-1)th* layer, $b_{ki}^l$ is the bias of the *kth* neuron in the lth layer, and ε is the learning rate.

Hyperparameters Setting

Five-fold cross-validation is employed to provide more confidence in the decision-making on the model structure to prevent overfitting.

First, the whole dataset is split into a test set on which the model performance will be evaluated, and a development set utilized to tune the CNN hyperparameters. The latter set is divided into five folds. Second, each fold is treated in turn as the validation set. In other words, the network is trained on the other four folds, and tested on the validation fold. Finally, the five validation losses are averaged to produce a more accurate error estimate since the model is tested on the full development set. The hyperparameters are then chosen such that they minimize the cross-validation error.

To test and compare the performance of these four gaze estimation techniques, a testing dataset was needed.

*3.3. Building a Database for Training and Testing*

All the discussed algorithms for gaze tracking require a database for the purpose of evaluation of the algorithms. Although three algorithms (template matching, LBP, and CNN) require a labeled dataset, the template matching algorithm that uses the eye pupil as a template does not require having such labels. There is more than one available database that could be used in testing the algorithms. Columbia Gaze Data Set [44] has 5880 images for 56 people in different head poses and eye gaze directions. Also, Gi4E [45] is another public dataset of iris center detection, it contains 1339 images that are collected using a standard webcam. UBIRIS [46] is another 1877-image database that is specified for iris recognition.

Nevertheless, neither of these databases was used because each one has one or more of the following drawbacks (with respect to our approach of testing).

1. The database may have changes in face position, which requires applying more stages to localize the eyes' area. Besides, the set-up for this project is based on having only one head pose, disregarding the gaze direction.
2. The database comprises only one gaze direction, which eliminates the possibility of using the dataset for testing for gaze tracking.
3. The dataset is not labeled, and the time and effort needed to label it is comparatively higher than building a similar new dataset.
4. One important feature that is missing in all of the available datasets is the transition between one gaze direction and another. This time should be known a priori, and be compared with the time needed by all the proposed algorithms.
5. Furthermore, all these datasets lack variations in lighting conditions.

Accordingly, a database was created for eight different users. Videos of eight users gazing in three directions (right, left, and forward) and closing their eyes were captured under two different lighting conditions. The dataset consists of the four selected classes, namely, Right, Forward, Left gazing, and one more class for closed eyes. For each user, a set of 2000 images were collected (500 for each class). The collected dataset includes images for indoors and outdoors lighting conditions. Besides, a dataset was collected for different ten users, yet, only for indoors lighting condition. Then, 500 frames per each class for each user were taken to constitute a benchmark dataset of 16,000 frames. Next, the dataset was partitioned into training set and testing set, where 80% of the data were used for training. Consequently, the algorithm learnt from 12,800 frames and were tested on 3200 unseen frames.

*3.4. Safety System—Ultrasonic Sensors*

For safety purposes, an automatic stopping mechanism installed on the wheelchair should disconnect the gazing-based controller, and then immediately stop the chair in case that the wheelchair becomes close to any object in the surroundings. Throughout this paper, the word "object" refers to any type of obstacle, which can be as small as a brick, or a chair, or even a human being.

A proximity sensor qualitatively measures how close an object is to the sensor. It has a specific range, and it raises a flag when any object enters this threshold area. There are two main technologies upon which proximity sensors operate: ultrasonic technology or infrared (IR) technology proximity sensors. In fact, the same physical concept applies for both, namely, wave reflections. The sensor transmits electromagnetic waves and receives them after they are reflected from surrounding objects. The process is very analogous to the operation of radar. The difference between ultrasonic-based and IR-based proximity sensors is obviously the type of the electromagnetic radiation.

The different types of proximity sensors can be used in different applications. As infrared sensors operate in the IR spectrum, they cannot be used in outdoors applications where sunlight interferes with their operation [47]. Besides, it is difficult to use IR sensors in dark areas [47]. On the other hand, ultrasonic waves are high-frequency sound waves that should not face any sort of interference. Furthermore, ultrasonic reflections are insensitive to hindrance factors as light, dust, and smoke. This makes ultrasonic sensors advantageous over IR sensors for the case of the wheelchair. A comparison between the two technologies [47] suggests that combining the two technologies together gives more reliable results for certain types of obstacles like rubber and cardboard. However, for this paper, there is no need to use the two sensors together; ultrasonic ones should be enough to accomplish the task of object detection.

Usually, an ultrasonic sensor is used along with a microcontroller (MCU) to measure the distance. As seen in Figure 10, the sensor has an ultrasonic transmitter (Tx) and a receiver (Rx). The microcontroller sends a trigger signal to the sensor, and this triggers the "Tx" to transmit ultrasonic waves. The ultrasonic waves then reflect from the object and are received by the Rx port on the sensor. The sensor accordingly outputs an echo signal (digital signal) whose length is equal to the time taken by the ultrasonic waves to travel (the double-way distance).

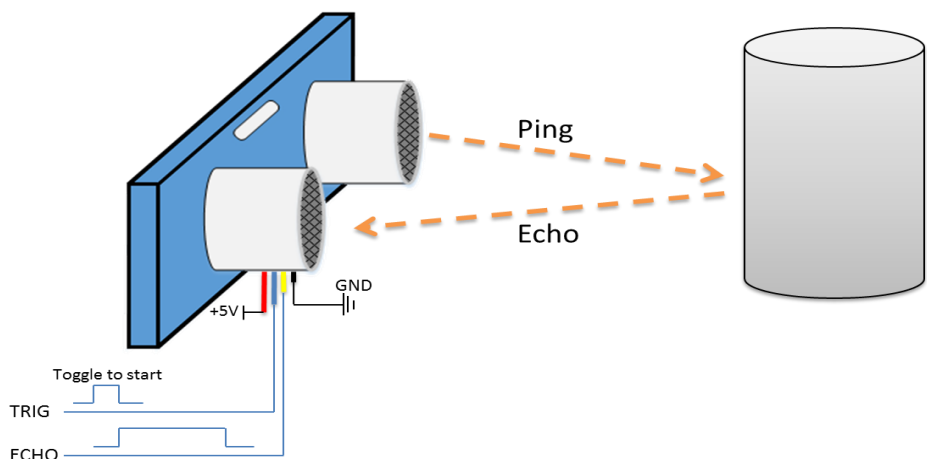

**Figure 10.** Proximity sensor with microcontroller working.

$$\text{distance}(m) = \frac{\text{travelling time}(\text{seconds})}{2} * \text{speed of ultrasonic waves}\left(\frac{\text{meter}}{\text{second}}\right) \quad (7)$$

$$\textbf{Error in distance} = \textbf{total delay in the system} * \textbf{speed of the wheelchair} \quad (8)$$

By knowing the travelling time of the signal, the distance can be calculated using Equation (7). Being analogous to the radars' operation, the time should be divided by two because the waves travel the distance twice. Ultrasonic waves travel in air with a speed of 340 m/sec.

As the wheelchair is moving with a certain speed, this affects the echo-based approach of measuring the distance (Doppler Effect). This introduces some inaccuracy in the calculated distance. However, this error did not affect the performance of the system. The first thing to do is to clarify that any fault in the measurement is completely dependent on the speed of the wheelchair, not the speed of the ultrasonic waves. This makes the error extremely small; as the error is directly proportional to the speed. The faulty distance can be calculated using Equation (8). The wheelchair moves at a maximum speed of 20 km/h (5.56 m/s). The delay is not a fixed parameter; thus, for design purposes, the maximum delay (worst-case scenario) was used.

The total delay in the system is the sum of delay introduced by the MCU and that of the sensor. The delay of the sensor strictly cannot exceed 1 millisecond because the average distance the sensor can measure is ~1.75 m. The average time of travelling of the ultrasonic wave can be calculated using Equation (7). The delay that may occur due to the MCU processing, it is in microseconds range. When applying this delay to Equation (8), the maximum inaccuracy in the measured distance is 0.011 m (1.1 cm).

To cover a wider range for the proximity sensors, a number of sensors was used to cover at 120⁰ as the chair is designed to move only in the forward direction (of course with right and left steering), but not towards back, refer Figure 11. A sketch program (Fusion 360) was used to visualize the scope of the proximity sensors and specify the number of sensors to use.

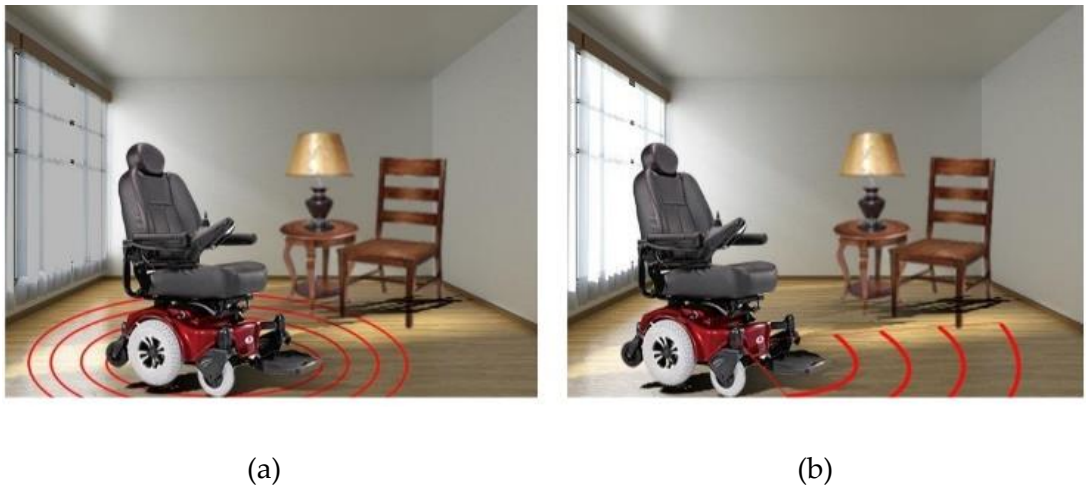

(a)          (b)

**Figure 11.** Proximity sensor covering range: (**a**) 360 degrees and (**b**) 120 degrees.

*3.5 Modifying the Wheelchair Controller*

The electric wheelchair Titan X16 has a joystick-based controller (DK-REMD11B). The objective is to adapt it for gaze direction control signals while maintaining the original joystick's functionality.

3.5.1. Joystick Control Mechanism

To tackle this problem, one must first uncover the underlying principle of operation of the built-in controller. After carrying out rigorous experimentation on the disassembled control unit, it was deduced that it operates according to the principle of electromagnetic induction.

As illustrated in Figure 12, the joystick handle is connected to a coil excited by a sinusoidal voltage and acts like a primary side of a transformer. Underneath the lever lies four coils in each direction (right, left, forward, and backward), which, although electrically isolated from the primary coil, are nevertheless magnetically coupled to it and they behave as the secondary side of a transformer. Consequently, a voltage was induced across each of the four secondary coils based on the rotation of the Joystick.

Furthermore, moving the joystick in a certain direction changes the coupling (i.e., mutual inductance) between them and the primary coil, thus altering the signals developed on each secondary coil. These signals were then processed to produce movement in the corresponding direction.

To preserve the joystick control capability, we devised a scheme that does not interfere with the built-in controller circuitry and was implemented on the existing joystick controller depicted in Figure 12. The designed system necessitates only three directions (right, left, and forward). There was mechanical arrangement was designed and implemented to control the existing joystick without modifying its functionality. The scheme for controlling the joystick is explained in details in the implementation section.

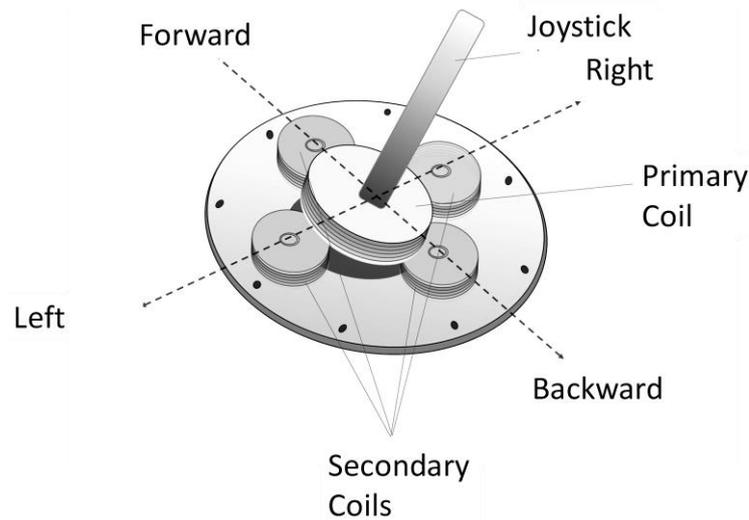

**Figure 12.** Working principle of existing joystick controller.

## 4. Implementation

This section describes the design methodology discussed in the previous section and implementation details of the final prototype. Implementation details cover eye frame design for image acquisition, Gaze estimation algorithm along with the dataset acquisition and training based on the dataset for the CNN, modified joystick controller, safety system implementation, and the final prototype.

*4.1. Frame Implementation*

The final frame is shown in Figure 5. The sunglasses were bought from a local grocery store, where the lenses were removed to not alter the wearers' vision. The white platforms are 3D printed to conveniently hold the IR cameras below eye level (Figure 5). The metallic arms holding the platforms were formed using clothing hangers, where the material provides enough flexibility to easily alter its position. The camera cables are tied to the sides of the sunglasses using zip-ties to make the overall unit more robust. The camera is equipped with a light-dependent resistor (LDR) to detect the amount of light present, which will automatically control the brightness of six on-board IR LEDs to illuminate accordingly. The system was checked under two extreme cases where there could be the possibility of deferring of the image quality due to absence and presence of strong sunlight. In small light variations, the IR LEDs brightness is automatically controlled by on-board LDR, and there were no significant image quality variations observed. The camera has a resolution of 2 megapixels and an adjustable lens focus. Also, it supports filming 1920 × 1080 at 30 frames per second, 1280 × 720 at 60 frames per second, and other filming configurations. Furthermore, the camera has a USB connection for computer interfacing.

*4.2. Gaze Estimation Algorithm*

In this paper, different solutions are proposed to estimate gaze directions: Template matching-based classifiers, LBP, and the CNN. Although the former alternative yielded a very satisfactory performance (i.e., average accuracy of 95%) for all indoors lighting condition, it failed to preserve this accuracy when images were collected outdoors because of the simplicity of the underlying algorithm. To clarify, this depreciation in performance is not because of the change of illumination, however, it is because of the physical difference of the shape of the eye between indoors and outdoors conditions. Two major changes occur when the user is exposed to direct sunlight: the iris shrinks when exposed to high intensity light and the eyelids tend to close.

One way to tackle this issue is to take distinct templates for the various lighting conditions that can be discerned by a light sensor, and then match the input with the template of the corresponding lighting condition. Conversely, the CNN is inherently complex enough to be able to generalize to all lighting conditions without the need of any additional hardware. Template matching technique was used as an aiding tool to the CNN to achieve better performance.

4.2.1. Collecting Training Dataset for the CNN—Calibration Phase

To make use of the accurate and fast CNNs, a training dataset should be available to train the model. The performance of the CNN is dependent on the selection of the training dataset; if the training dataset is not properly selected, the accuracy of the gaze estimation system dramatically decreases. Thus, a calibration phase is required for any new user of the wheelchair. The output of this calibration phase is the trained CNN model that can be used later for gaze estimation. To prepare a nice training dataset, a template matching technique was used. Template matching does not require any prior knowledge of the dataset. Note that the calibration phase faces two problems: the first problem arises from the fact that the user blinks during the collection of the training data. This leads to a flawed model prediction. The second issue is that the user may not respond immediately to the given instructions, i.e., the user may be asked to gaze right for couple of seconds, but he or she responds one or two seconds later.

Fortunately, template matching, if smartly used, can overcome these two problems, and can be used to make sure that the training dataset is 100% correctly labeled. Moreover, the training dataset should be diverse enough to account for different lighting conditions, and different placement of glasses. The cameras could zoom the user's eyes, thus if the user slightly changes the position of glasses, noticeable differences occur between the images.

Figure 13 shows a flowchart of the calibration phase. For simplicity, illustration is discussed on only one eye; however, the same applies for the second eye. First, to have a wide-ranging dataset, the calibration stage was repeated for different conditions, each condition is named as a Scenario.

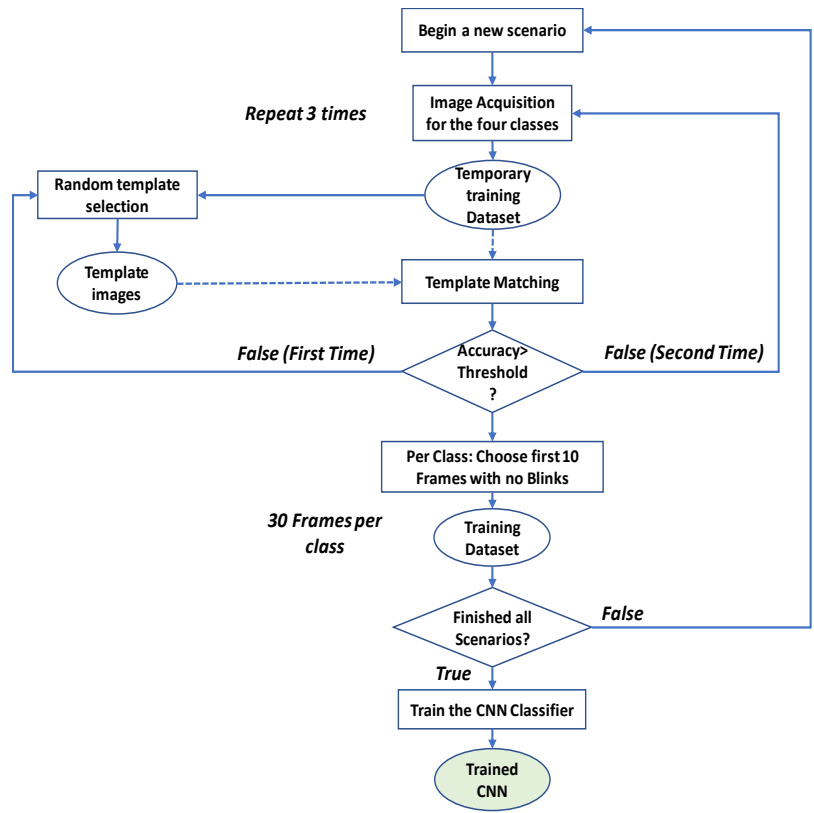

**Figure 13.** Flowchart for the calibration phase.

For each lighting condition, the user was asked to change the position of glasses (slightly slide the glasses either down or up). Thus, four different scenarios were needed. However, that is not enough to ensure a diverse dataset. Therefore, for each scenario, images were acquired for the four classes in three nonconsecutive attempts to capture all the possible patterns of the user's gaze in a particular direction. Every time, the user, indeed, gazed in a different manner. Thus, the importance of collecting images three times can be seen. Each time, 200 frames were collected (a net of 600 frames per class). These frames (that are contaminated with blinks) constitute a temporary training dataset. As previously mentioned, template matching was used to remove these blinks; a random frame was selected from each class to be used as a template.

There are numerous scenarios that may happen: the first and most probable scenario is that the user has followed the instructions and the template selection was successful, that is, the template image is not a blinking frame. In such a case, and from the testing on the benchmark dataset, the accuracy of template matching was higher than 80%. This accuracy had been set as the threshold; if this accuracy was attained, this means that the image acquisition was successful. The second possibility was that while running the template-matching algorithm, a low accuracy was returned. The cause of this low accuracy can be wrong acquisition or wrong template selection. Thus, it was better to test first for a wrong template selection; thus, a different template was randomly selected, and the template matching was carried out again. If the accuracy increases to reach the threshold, the issue was with the template; i.e., it might have been an irrelevant blinking image. However, if the low accuracy persists, there is a high probability that the user has gazed in a wrong direction during the image acquisition phase, and therefore the image acquisition was repeated.

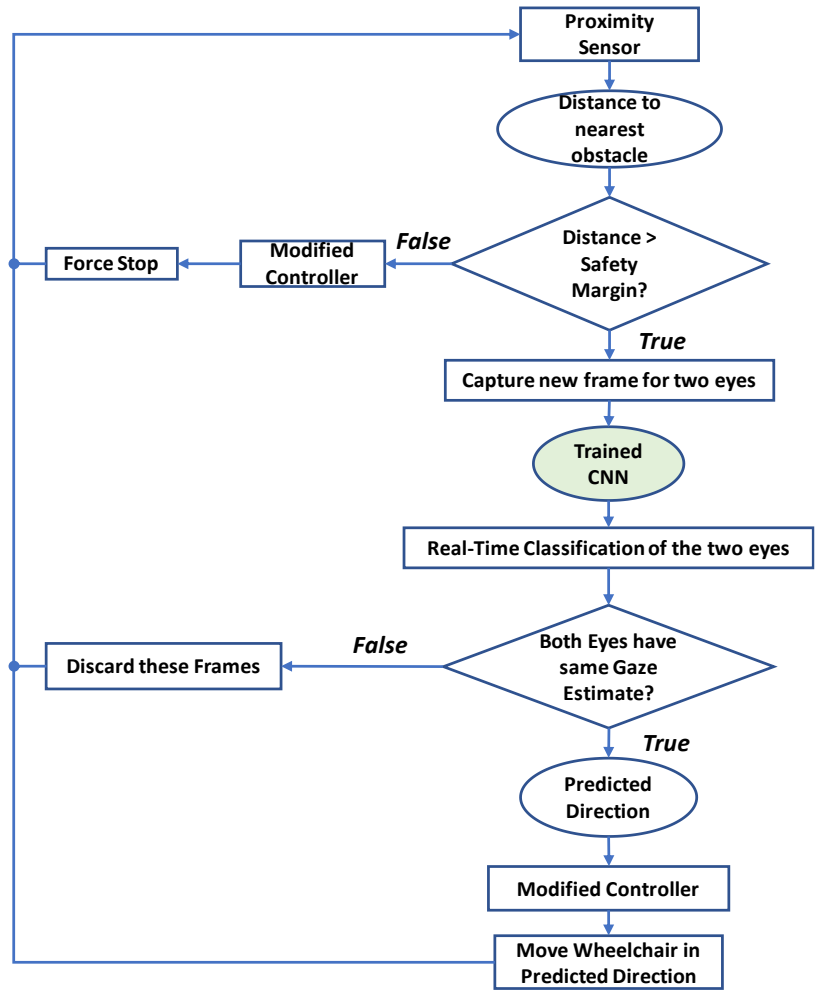

**Figure 14.** Flowchart for real-time classification and controlling the movement of the wheelchair.

With the threshold accuracy obtained, it was certainly known that the template image is correctly selected. The next stage was to clean the 600 images of each class from those blinking frames and to keep 500 useful images for each class. Then, 400 images were randomly selected for training while 100 images tested the trained model. After finishing all the four scenarios, the final training dataset was fed to the CNN to get the trained CNN model. It was worth mentioning that even if this calibration requires few minutes to accomplish, it is required to execute once only.

By having the trained CNN, the system was ready for real-time gaze estimation to steer the wheelchair. Between the real-time gaze estimation and moving the wheelchair, there are two points that should be taken into consideration. The first is the safety of the user; whether there is an obstacle in his/her way. The second point is that the real-time estimation for the right eye may give a different classification from the left eye (because of a wrong classification or because of different gazing at the instance the images are collected).

Figure 14 shows a flowchart for the real-time classification, where these two points are tackled. As discussed earlier, proximity sensors were used to scan the region in the wheelchair way. In the decision-making part, the minicomputer first measured the distance for the closest obstacle. If there was no close obstacle, then the wheelchair moved in the predicted direction. Nevertheless, if there was an object in the danger area (threshold), the wheelchair stopped immediately.

If there was no close obstacle, when the real-time classification runs, two classification results are returned, one for each eye. In fact, it is very important to make use of this redundancy of results; otherwise, there is no need of having two cameras. If the two eyes returned the same class, and considering the high classification performance of the CNN, this predicted class was used to move the wheelchair in the determined direction. However, if the two eyes gave different classifications, no action will be taken, as the safety of the user had the priority.

4.2.2. Training the CNN

The 2-D CNN-based gaze estimator had a relatively shallow and compact structure with only two CNN layers, two subsampling layers, and two fully connected (FC) layers, as illustrated in Table 1. This boosted the system's computational efficiency for training, and most importantly for real-time classification. In this configuration, the subsampling factor (in both dimensions) of the last subsampling layer was adaptively set to 13 to ensure that its output was a stack of 1 × 1 feature maps (i.e. scalar features).

**Table 1.** CNN hyperparameters.

| CNN Layers | | | | Subsampling Layers | | | FC Layers | Activation Function |
|---|---|---|---|---|---|---|---|---|
| Number | Filter Size | Stride | Zero Padding | Number | Pooling Type | Subsampling Factor (x,y) | Number | |
| 2 | 3 | 1 | 0 | 2 | max | 4 | 2 | tanh |

The used network had 16 and 12 filters (i.e., hidden neurons) in the two hidden CNN layers, respectively, and 16 hidden neurons in the hidden FC layer. The output layer size was 4, which corresponds to the number of classes. RGB images were fed to the 2-D CNN, and thus the input depth was 3 channels, each of which was a 64 × 64 frame. Training was conducted by means of backpropagation (BP) with 3 stopping criteria: the minimum train classification error was 1%, the maximum number of BP iterations was 100, or the minimum loss gradient was 0.001 (optimizer converges).

The learning rate ε was initialized to 0.001 and then global adaptation was carried out for each BP iteration: ε was increased by 5% in the next iteration if the training loss decrease, and it was reduced by 30% otherwise.

*4.3. Modifying the Joystick Controller*

The original controller was kept intact to maintain its original functions, while external modifications were done to the joystick of the controller as shown in Figure 15. The gaze estimation algorithm produces decision on every frame; however, the average of each 10 frames was used to send command to the joystick. This step reduces any potential random error of the control system and reduces ambiguous command. An Arduino Uno microcontroller board was interfaced to Mini-computer using USB interface, which received the commands from gaze algorithm and controlled two servo motors accordingly. Figure 16 shows how two servo motors can produce stop, forward, left, and right directions using two slider with the help of existing joystick controller.

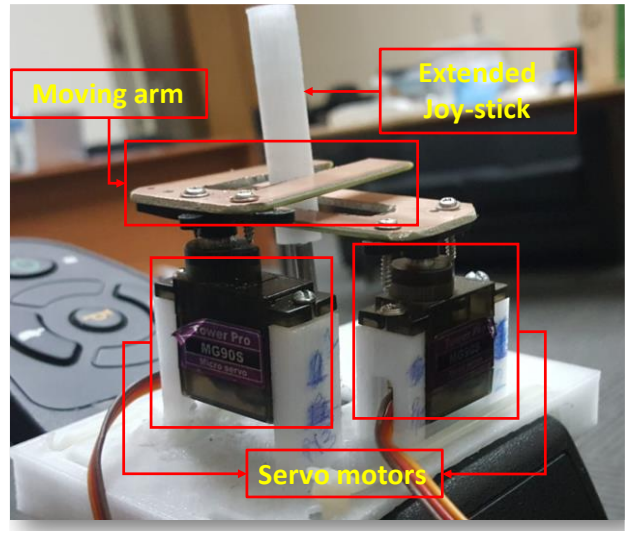

**Figure 15.** Two servo motors along with two moving arms were attached to original joystick with 3D printed support structure for motion control.

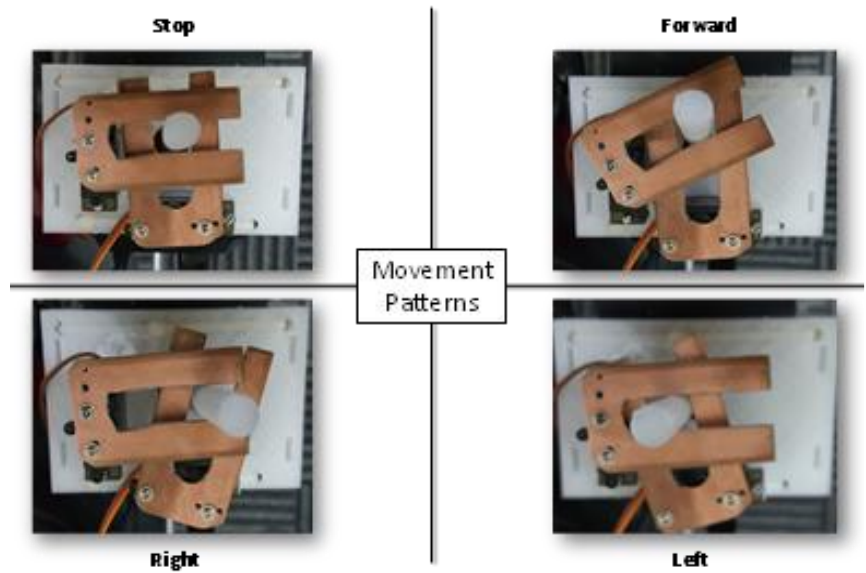

**Figure 16.** Four different positions of the servo motors to control four different classes of the system.

*4.4. Safety System Implementation*

As discussed in the methodology section, an array of ultrasonic sensors was needed to cover the all mobility range of the wheelchair. Two assumptions were made when choosing the number and positions of proximity sensors: the stopping distance is 1 m away from detected objects and that the electric wheelchair stops instantaneously. The electric wheelchair moves only forwards and rotates. Thus, the sensors were placed in front of the wheelchair to detect outward objects. An aluminum frame was built and mounted on the wheelchair to hold the sensors.

The three outward sensors, as shown in Figure 17, provided a horizontal detection range of 1 meter, with gaps of 8 centimeters between the covered regions at a 1 m distance from the aluminum frame. Moreover, the three sensors provided a 43 degrees detection angle. Two sensors were placed in between the three previous sensors, but slanted downwards. These sensors detected obstacles shorter than the aluminum frame.

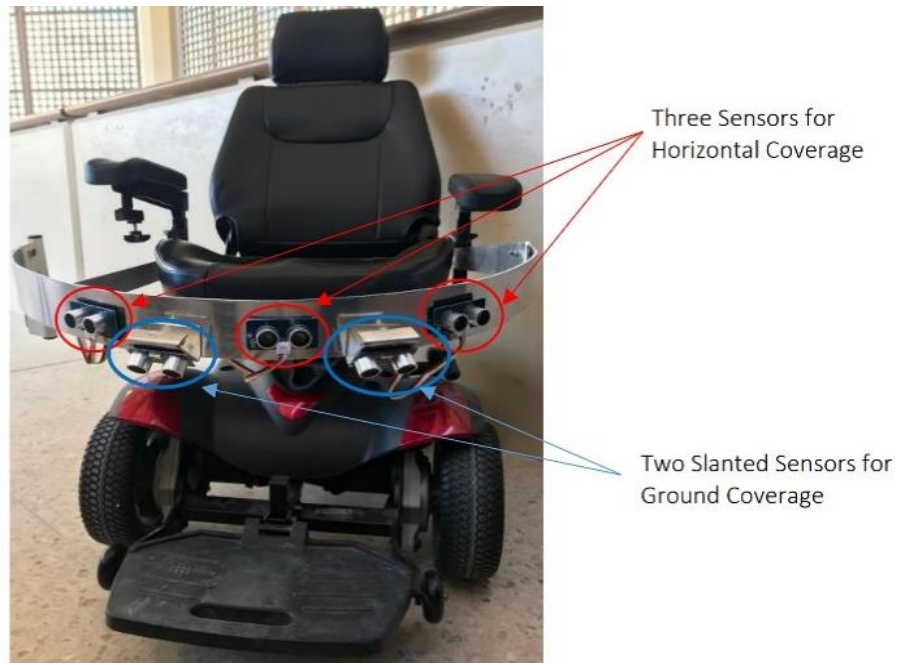

**Figure 17.** Implemented proximity sensor design.

## 5. Results and Discussion

The complete system prototype is presented in Figure 17. This section provides the details of the metrics which can show the performance of real time wheel chair movement for disabled people using eye gazing.

*5.1. Computation Complexity Analysis*

The average time for one backpropagation (BP) iteration per frame is ~19.7 ms with the aforementioned computer implementation. Considering a full BP run with the maximum of 100 iterations over the train set of 60 (for examples), the maximum training time is 60 × 100 × 19.7 = 1.97 min. The total time for a forward pass of a single image to obtain the score vector, which corresponds to the system latency is about 1.57 ms. A system with a latency of 50 ms or less, is considered a real-time system. However, the proposed system operates at a speed that is higher than 30x the real-time speed. The average frame rate that was practically achieved was ~ 99 frames per sec (fps). This is a significant increase in the frame rate comparing other works reported in the literature.

*5.2. Real-time Performance*

A Softmax layer was added after the CNN in the real-time gaze estimation. As this is a four-class classification problem, the Softmax block mapped the 4-D class scores vector, output by the CNN, to an equivalent 4-D class probabilities vector. The probabilities in this vector quantify the model's confidence level in thinking that the input frame is of a particular class. Ultimately, the one with the highest probability value was the predicted class. A sample of the results is shown in Figure 18.

Table 2. Normalized confusion matrix of % of classification results.

|  |  | Actual | | | |
|---|---|---|---|---|---|
|  |  | Right | Forward | Left | Closed |
| Predicted | Right | 98.75 | 0 | 1.25 | 0 |
|  | Left | 1.56 | 98.44 | 0 | 0 |
|  | Forward | 0 | 0 | 100 | 0 |
|  | Closed | 0 | 0 | 0 | 100 |

```
D:\Users\dahmani\Desktop\QU\DAHH\ML\CNN\CNN F
41% 13% 33% 13% Right
42% 13% 33% 13% Right
41% 13% 34% 13% Right
41% 13% 33% 12% Right
41% 13% 33% 12% Right
42% 13% 32% 13% Right
41% 14% 32% 13% Right
42% 13% 33% 13% Right
41% 13% 34% 12% Right
42% 13% 33% 12% Right
43% 13% 32% 12% Right
42% 13% 32% 13% Right
43% 12% 33% 12% Right
41% 13% 34% 12% Right
39% 12% 38% 11% Right
36% 14% 37% 13% Left
36% 14% 37% 13% Left
34% 15% 38% 14% Left
34% 15% 37% 14% Left
33% 15% 38% 14% Left
33% 15% 38% 14% Left
33% 15% 38% 14% Left
33% 15% 38% 15% Left
32% 15% 38% 14% Left
33% 14% 38% 14% Left
33% 15% 37% 14% Left
34% 14% 38% 14% Left
36% 13% 39% 12% Left
36% 13% 39% 12% Left
```

Figure 18. Real time prediction results.

*5.3. Classification Results*

Five-fold cross-validation was performed to obtain precise classification accuracy (i.e., the ratio of the number of correctly classified patterns to the total number of patterns classified). This estimates the model's true ability to generalize and extrapolate to new unseen data. This was accomplished by testing on 20% of the entire benchmark dataset of eight users that contains 3200 images. With 5-fold cross-validation, five CNNs were trained, and therefore five 4x4 confusion matrices (CM) per user were computed. These five CMs were then accumulated to yield a cross-validation confusion matrix per user, out of which the cross-validation classification accuracy for every user was computed and shown in Figure 19. Subsequently, the eight cross-validation CMs were accumulated to produce an overall CM that represents the complete model performance over the whole dataset as illustrated in Table 2. Finally, the overall CNN classification accuracy was computed. It is worth noting from Figure 19 that the lowest cross-validation classification accuracy among all the subjects is 96.875%, which is quite satisfactory for accurately estimating a user's gaze. Moreover, five out eight users had a 100% cross-validation accuracy.

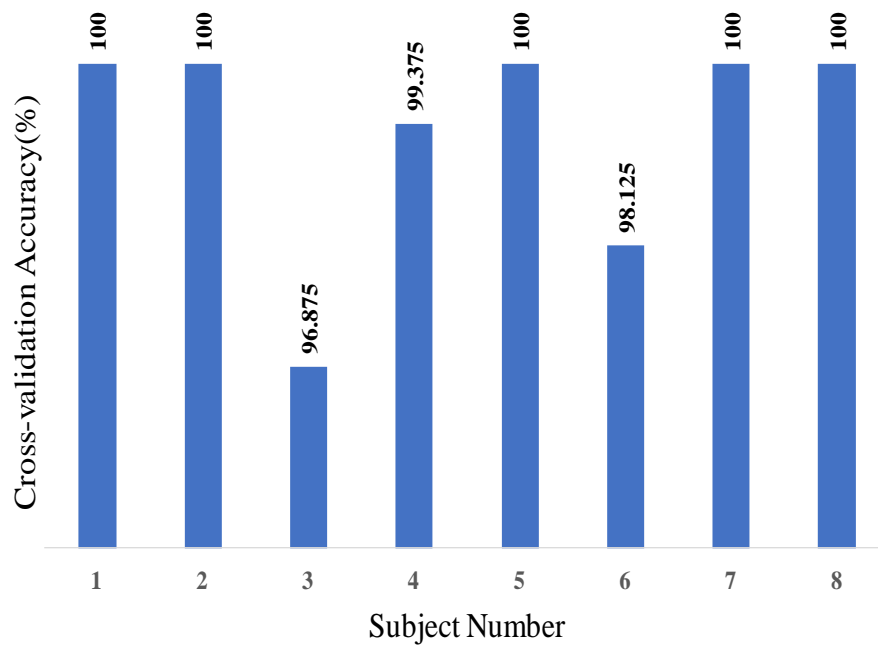

**Figure 19.** Cross-validation accuracies of eight subjects.

According to the normalized CM shown in Table 2, the classification result of the designed CNN yields probabilities higher than 98.4% for an accurate gaze estimation for all classes. As the ground truth probabilities (columns) indicate, right and forward classes have extremely low confusion probabilities (1.25% and 1.56%) to left and forward classes, respectively. The overall CNN classification accuracy is 99.3%, which makes the gaze estimation reliability of the designed system practically 100%. The implemented system's performance is summarized in Table 3.

**Table 3.** Summary of the implemented system's evaluation metrics.

| | |
|---|---|
| Accuracy (%) | 99.3 |
| Frame Rate (frames/sec) | 99 |
| Maximum Training Time (min) | 1.97 |

**6. Conclusions**

This paper aimed to design and implement a motorized wheelchair that can be controlled via eye movements. This is considered as a pivotal solution for people with complete paralysis of four limbs. The starting point for this project was a joystick-controlled wheelchair; all other building blocks were designed from scratch. Hardware-wise, the design stage aimed to build an image acquisition system and to modify the existing controller to move the wheelchair based on commands given from users' eyes instead of the manual joystick. The core and most challenging part of this prototype was to build a gaze estimation algorithm. Two approaches were followed: machine learning-based algorithms and template matching-based ones. Both approaches showed extremely high accuracy for a certain lighting condition. Nevertheless, template matching did not have a robust performance, as the accuracy had dropped dramatically when it was tested in another lighting condition. The Convolutional Neural Networks had superior performance in this regard, and were therefore the chosen technique for gaze estimation. Based on Table 3, the proposed system is very suitable for real time application. The classification accuracy of the gaze estimation is 99.3% which is above 95%, i.e., the norm for such type of applications. The model can process 99 frames per second and can make a prediction in ~1.57 ms, which 5x faster than the real-time requirements and the recent works reported in the literature. The safety of the user was given a paramount concern, and was ensured by incorporating an array of five ultrasonic sensors to cover the whole range of motion of the wheelchair. It can instantly stop if any obstacle is detected in this range. The system can be further developed by opting for a global approach rather than a user specific one, to eliminate the need for a calibration phase. Moreover, the current gaze estimation algorithm can stop the wheelchair or make it move only in three directions, namely, forward, right, and left. However, if a regression was used instead of classification, more degrees of freedom could be provided for movement. The authors are investigating on the joystick controller protocol to replicate it in the microcontroller used for joystick control. In this way, a non-mechanical system of a joystick controller can be developed to control the wheelchair controller directly.

**Author Contribution:** Experiments were designed by M.E.H.C. and S.K. Experiments were performed by M.D., T.R., K.A.J., and A.H. Results were analyzed by M.D., M.E.H.C. and S.K. All authors were involved in the interpretation of data and paper writing.

**Funding:** The publication of this article was funded by the Qatar National Library and Qatar National Research Foundation (QNRF), grant numbers NPRP12S-0227-190164 and UREP22-043-2-015.

**Conflicts of Interest:** The authors declare no conflicts of interest.

**References**

1. Analysis of Wheelchair Need/Del-Corazon.org. Available online: http://www.del-corazon.org/analysis-of-wheelchair-need (accessed on 10 March 2017).
2. Zandi, A.S.; Quddus, A.; Prest, L.; Comeau, F.J.E. Non-Intrusive Detection of Drowsy Driving Based on Eye Tracking Data. *Transp. Res. Rec. J. Transp. Res. Board* 2019, *2673*, 247–257.
3. Zhang, J.; Yang, Z.; Deng, H.; Yu, H.; Ma, M.; Xiang, Z. Dynamic Visual Measurement of Driver Eye Movements. *Sensors* 2019, *19*, 2217.
4. Strobl, M.A.R.; Lipsmeier, F.; Demenescu, L.R.; Gossens, C.; Lindemann, M.; De Vos, M. Look me in the eye: Evaluating the accuracy of smartphone-based eye tracking for potential application in autism spectrum disorder research. *Biomed. Eng. Online* 2019, *18*, 51.
5. Shishido, E.; Ogawa, S.; Miyata, S.; Yamamoto, M.; Inada, T.; Ozaki, N. Application of eye trackers for understanding mental disorders: Cases for schizophrenia and autism spectrum disorder. *Neuropsychopharmacol. Rep.* 2019, *39*, 72–77.
6. Cruz, R.; Souza, V.; Filho, T.B.; Lucena, V. Electric Powered Wheelchair Command by Information Fusion from Eye Tracking and BCI. In Proceedings of the 2019 IEEE International Conference on Consumer Electronics (ICCE), Las Vegas, NV, USA, 11–13 January *2019*.
7. Rupanagudi, S.R.; Koppisetti, M.; Satyananda, V.; Bhat, V.G.; Gurikar, S.K.; Koundinya, S.P.; Sumedh., S.K.M.; Shreyas, R.; Shilpa, S.; Suman, N.M.; et al. A Video Processing Based Eye Gaze Recognition Algorithm for Wheelchair Control. In Proceedings of the 2019 10th International Conference on Dependable Systems, Services and Technologies (DESSERT), Leeds, UK, 5–7 June *2019*.
8. Kumar, A.; Netzel, R.; Burch, M.; Weiskopf, D.; Mueller, K. Visual Multi-Metric Grouping of Eye-Tracking Data. *J. Eye Mov. Res.* **2018**, *10*, 17.
9. Ahmed, H.M.; Abdullah, S.H.; A Survey on Human Eye-Gaze Tracking (EGT) System "A Comparative Study". *Iraqi J. Inf. Technol.* 2019, *9*, 177–190.
10. Vidal, M.; Turner, J.; Bulling, A.; Gellersen, H. Wearable eye tracking for mental health monitoring. *Comput. Commun.* 2012, *35*, 1306–1311.
11. Reddy, T.K.; Gupta, V.; Behera, L. Autoencoding Convolutional Representations for Real-Time Eye-Gaze Detection. In *Advances in Intelligent Systems and Computing*; Springer Science and Business Media LLC: Singapore, Singapore, 2018; pp. 229–238.

12. Jafar, F.; Fatima, S.F.; Mushtaq, H.R.; Khan, S.; Rasheed, A.; Sadaf, M. Eye Controlled Wheelchair Using Transfer Learning. In Proceedings of the 2019 International Symposium on Recent Advances in Electrical Engineering (RAEE), Islamabad, Pakistan, 28–29 August 2019.
13. Deshpande, S.; Adhikary, S.D.; Arvindekar, S.; Jadhav, S.S.; Rathod, B. Eye Monitored Wheelchair Control for People Suffering from Quadriplegia. *Univers. Rev.* 2019, *8*, 141–145.
14. Majaranta, P.; Bulling, A. Eye Tracking and Eye-Based Human–Computer Interaction. In *Human–Computer Interaction Series*; Springer Science and Business Media LLC: London, United Kingdom, 2014; pp. 39–65.
15. Hickson, S.; Dufour, N.; Sud, A.; Kwatra, V.; Essa, I. Eyemotion: Classifying Facial Expressions in VR Using Eye-Tracking Cameras. In Proceedings of the 2019 IEEE Winter Conference on Applications of Computer Vision (WACV), Waikoloa Village, HI, USA, 7–11 January 2019.
16. Harezlak, K.; Kasprowski, P. Application of eye tracking in medicine: A survey, research issues and challenges. *Comput. Med. Imaging Graph.* 2018, *65*, 176–190.
17. Fuhl, W.; Tonsen, M.; Bulling, A.; Kasneci, E. Pupil detection for head-mounted eye tracking in the wild: An evaluation of the state of the art. *Mach. Vis. Appl.* 2016, *27*, 1275–1288.
18. Liu, T.-L.; Fan, C.-P. Visible-light wearable eye gaze tracking by gradients-based eye center location and head movement compensation with IMU. In Proceedings of the 2018 IEEE International Conference on Consumer Electronics (ICCE), Las Vegas, NV, USA, 12–14 January *2018*.
19. Robbins, S.; McEldowney, S.; Lou, X.; Nister, D.; Steedly, D.; Miller, Q.S.C.; Bohn, D.D.; Terrell, J.P.; Goris, A.C.; Ackerman, N. Eye-Tracking System Using a Freeform Prism and Gaze-Detection Light. U.S. Patent 10,228,561, 12 March 2019.
20. Sasaki, M.; Nagamatsu, T.; Takemura, K. Screen corner detection using polarization camera for cross-ratio based gaze estimation. In Proceedings of the 11th ACM Symposium on Eye Tracking Research & Applications, Denver, CO, USA, 25–28 June 2019.
21. Holland, J. Eye Tracking: Biometric Evaluations of Instructional Materials for Improved Learning. *Int. J Educ. Pedag. Sci.* 2019, *13*, 1001–1008.
22. Chen, B.-C.; Wu, P.-C.; Chien, S.-Y. Real-time eye localization, blink detection, and gaze estimation system without infrared illumination. In Proceedings of the 2015 IEEE International Conference on Image Processing (ICIP), Quebec, QC, Canada, 27–30 September 2015.
23. Fuhl, W.; Santini, T.C.; Kübler, T.; Kasneci, E. Else: Ellipse selection for robust pupil detection in real-world environments. In Proceedings of the Ninth Biennial ACM Symposium on Eye Tracking Research & Applications, Charleston, SC, USA, 14–17 March 2016.
24. Fuhl, W.; Kübler, T.; Sippel, K.; Rosenstiel, W.; Kasneci, E. ExCuSe: Robust Pupil Detection in Real-World Scenarios. In *International Conference on Computer Analysis of Images and Patterns*; Springer: Cham, Germany 2015; pp. 39–51.
25. Kassner, M.; Patera, W.; Bulling, A. Pupil: An Open Source Platform for Pervasive Eye Tracking and Mobile Gaze-based Interaction. In Proceedings of the 2014 ACM International Joint Conference on Pervasive and Ubiquitous Computing: Adjunct Publication, Seattle, WA, USA, September 2014.
26. Javadi, A.-H.; Hakimi, Z.; Barati, M.; Walsh, V.; Tcheang, L. SET: A pupil detection method using sinusoidal approximation. *Front. Neuroeng.* 2015, *8*, 4.
27. Li, D.; Winfield, D.; Parkhurst, D.J. Starburst: A hybrid algorithm for video-based eye tracking combining feature-based and model-based approaches. In Proceedings of the 2005 IEEE Computer Society Conference on Computer Vision and Pattern Recognition (CVPR'05)-Workshops, San Diego, CA, USA, 21–23 September 2005.
28. Świrski, L.; Bulling, A.; Dodgson, N. Robust real-time pupil tracking in highly off-axis images. In Proceedings of the Symposium on Eye Tracking Research and Applications, Santa Barbara, CA, USA, 28–30 March 2012.
29. Mompeán, J.; Aragon, J.L.; Prieto, P.M.; Artal, P. Design of an accurate and high-speed binocular pupil tracking system based on GPGPUs. *J. Supercomput.* 2017, *74*, 1836–1862.
30. Naeem, A.; Qadar, A.; Safdar, W. Voice controlled intelligent wheelchair using raspberry pi. *Int. J. Technol. Res.* 2014, *2*, 65.
31. Rabhi, Y.; Mrabet, M.; Fnaiech, F. A facial expression controlled wheelchair for people with disabilities. *Comput. Methods Programs Biomed.* 2018, *165*, 89–105.
32. Arai, K.; Mardiyanto, R.; Nopember, K.I.T.S. A Prototype of ElectricWheelchair Controlled by Eye-Only for Paralyzed User. *J. Robot. Mechatron.* 2011, *23*, 66–74.

33. Mani, N.; Sebastian, A.; Paul, A.M.; Chacko, A.; Raghunath, A. Eye controlled electric wheel chair. *Int. J. Adv. Res. Electr. Electron. Instrum. Eng.* 2015, *4*, doi:10.15662/ijareeie.2015.0404105.
34. Gautam, G.; Sumanth, G.; Karthikeyan, K.; Sundar, S.; Venkataraman, D. Eye movement based electronic wheel chair for physically challenged persons. *Int. J. Sci. Technol. Res.* 2014, *3*, 206–212.
35. Patel, S.N.; Prakash, V.; Narayan, P.S. Autonomous camera based eye controlled wheelchair system using raspberry-pi. In Proceedings of the 2015 International Conference on Innovations in Information, Embedded and Communication Systems (ICIIECS), Coimbatore, India, 19–20 March 2015.
36. Chacko, J.K.; Oommen, D.; Mathew, K.K.; Sunny, N.; Babu,N. Microcontroller based EOG guided wheelchair. *Int. J. Med. Health Pharm. Biomed. Eng.* 2013, *7*, 409–412.
37. Al-Haddad, A.; Sudirman, R.; Omar, C. Guiding Wheelchair Motion Based on EOG Signals Using Tangent Bug Algorithm. In Proceedings of the 2011 Third International Conference on Computational Intelligence, Modelling & Simulation, Langkawi, Malaysia, 20–22 September 2011.
38. Elliott, M.A.; Malvar, H.; Maassel, L.L.; Campbell, J.; Kulkarni, H.; Spiridonova, I.; Sophy, N.; Beavers, J.; Paradiso, A.; Needham, C.; et al. Eye-controlled, power wheelchair performs well for ALS patients. *Muscle Nerve* 2019, *60*, 513–519.
39. Krafka, K.; Khosla, A.; Kellnhofer, P.; Kannan, H.; Bhandarkar, S.; Matusik, W.; Torralba, A. Eye Tracking for Everyone. In Proceedings of the 2016 IEEE Conference on Computer Vision and Pattern Recognition, Las Vegas, NV, USA, 26 June–1 July 2016.
40. George, A.; Routray, A. Real-time eye gaze direction classification using convolutional neural network. In Proceedings of the 2016 International Conference on Signal Processing and Communications (SPCOM), Bangalore, India, 12–15 June 2016.
41. Spinel IR Camera, Spinel 2MP full HD USB Camera Module Infrared OV2710 with Non-distortion Lens FOV 100 degree, Support 1920x1080@30fps, UVC Compliant, Support most OS, Focus Adjustable UC20MPD_ND. Available online: https://www.amazon.com/Spinel-Non-distortion-1920x1080-Adjustable-UC20MPD_ND/dp/B0711JVGTN (accessed on10 March 2017).
42. He, D.-C.; Wang, L. Texture Unit, Texture Spectrum, and Texture Analysis. *IEEE Trans. Geosci. Remote. Sens.* 1990, *28*, 509–512.
43. Lian, Z.; Er, M.J.; Li, J. A Novel Face Recognition Approach under Illumination Variations Based on Local Binary Pattern. In Proceedings of the International Conference on Computer Analysis of Images and Patterns, Seville, Spain, 29–31 August 2011.
44. Cs.columbia.edu. CAVE/Database: Columbia Gaze Data Set. 2017. Available online: http://www.cs.columbia.edu/CAVE/databases/columbia_gaze/ (accessed on10 March 2017).
45. GI4E/Gi4E Database. Available online: Ttp://gi4e.unavarra.es/databases/gi4e/ (accessed on 10 March 2017).
46. UBIRIS Database. Available online: http://iris.di.ubi.pt/ (accessed on 10 March 2017).
47. Lakovic, N.; Brkic, M.; Batinic, B.; Bajic, J.; Rajs, V.; Kulundzic, N. Application of low-cost VL53L0X ToF sensor for robot environment detection. In Proceedings of the 2019 18th International Symposium Infoteh-Jahorina (INFOTEH), East Sarajevo, Srpska, 20–22 March 2019.